\documentclass[journal,onecolumn,12pt]{IEEEtran}

\usepackage{amssymb,amsmath,epsfig,graphicx,theorem}
\usepackage{amsfonts}
\usepackage{bm}
\usepackage{arydshln}
\usepackage[normalem]{ulem}
\usepackage[linesnumbered,ruled,vlined]{algorithm2e}
\newtheorem{Theo}{Theorem}
\newtheorem{Lem}{Lemma}

\newtheorem{Remark}{Remark}

\SetKwInput{KwInput}{Input}                
\SetKwInput{KwOutput}{Output}              
\SetKwInput{Initialization}{Initialization}  

\usepackage{epsfig,rotating,setspace,latexsym,amsmath,epsf,amssymb,bm,color}
\usepackage{cite}

\IEEEoverridecommandlockouts
\allowdisplaybreaks[4]

\title{On the Asymptotic Capacity of Information Theoretical Privacy-preserving Epidemiological Data Collection}

\author{Jiale Cheng, Nan Liu, and Wei Kang
\thanks{J. Cheng and N. Liu are with the National Mobile Communications Research Laboratory,
Southeast University, Nanjing, China (email: \{jlcheng, nanliu\}@seu.edu.cn). 
W. Kang is with the School of Information Science and Engineering,
Southeast University, Nanjing, China (email: wkang@seu.edu.cn). }
}

\begin{document}

\maketitle
\begin{abstract}

We formulate a new secure distributed computation problem, where a simulation center can require any linear combination of $ K $ users' data through a caching layer consisting of $ N $ servers. The users, servers, and data collector do not trust each other. For users, any data is required to be protected from up to $ E $ servers; for servers, any more information than the desired linear combination cannot be leaked to the data collector; and for the data collector, any single server knows nothing about the coefficients of the linear combination. Our goal is to find the optimal download cost, which is defined as the size of message uploaded to the simulation center by the servers, to the size of desired linear combination. We proposed a scheme with the optimal download cost when $E < N-1$. We also prove that when $E\geq N-1$, the scheme is not feasible.
\end{abstract}
\begin{IEEEkeywords}
secure multiparty computation, epidemiological data collection, asymptotic capacity
\end{IEEEkeywords}

\section{Introduction}
During any prevention and control period in the epidemic, strengthening the protection of personal information is conducive not only to safeguarding personal interests, but also better controlling the development of the epidemic. In epidemiological modeling, many recent studies have shown that various models have a good fitting effect on the nature of the epidemic, such as the Bayesian model\cite{Anderson2020QuantifyingTI}, and the deep learning model including multi-head attention, long short-term memory (LSTM), and convolutional neural network (CNN)\cite{ABBASIMEHR2021110511}. However, the simulation process still can not get rid of the strong dependence on personal data. At the same time, the model adjustment required for a large number of personal data is also a technical problem that needs to be solved urgently. Epidemiological modeling generally requires a collection of different types of information uploaded by users in a real-time way, but in fact, data collection does not require all the details of users to quantitatively analyze the epidemiological nature. The solution to the contradiction between data analysis and data protection leads us to the theoretical analysis of the privacy-preserving epidemiological data collection problem.

In modeling epidemiological data collection, a large and changing number of users submit their physical data to an untrusted server at a specified time. Additionally, the data collector who conducts epidemiological modeling retrieves the corresponding data by accessing the server. In real contact graphs where physical data is collected by mobile devices, the analytic data of users in certain regions are essential for epidemiological study to get a proper estimation of the potential public health hazards. Unlike user uploads, these data collectors are only interested in some statistical features contained in the data stored by the server. It is worth noting that users, servers, and data collectors do not trust each other; that is, users need to ensure that their data is confidential to the server, and data collectors cannot know the details of any single user. At the same time, data collectors do not want the server to know the characteristics of the users' data they are interested in.

In the analysis of communicable diseases including COVID-19, a detailed model with sufficient interaction data is required\cite{Daniel2021PEM}. Concerning the security hazard and privacy leakage of data, Several studies in information theory have focused on the issue when sharing messages to untrusted agencies \cite{Wan2021DLSC,Wan2021TradeoffDLSC,Wan2022SecureDLSC,chen2020gcsa,zhao2021SA,Chang2018DMM,Has2022BPC,Netanel2020CyclicMDSGC}. In this study, we present a practical framework for privacy-preserving epidemiological data collection problem and analyze the capability that a data collectors can receive the shared data securely and privately, with respect to the number of symbols they need to download.

\section{System Model}
\begin{figure}[htbp]
	\centering
	\includegraphics[width=0.7\linewidth]{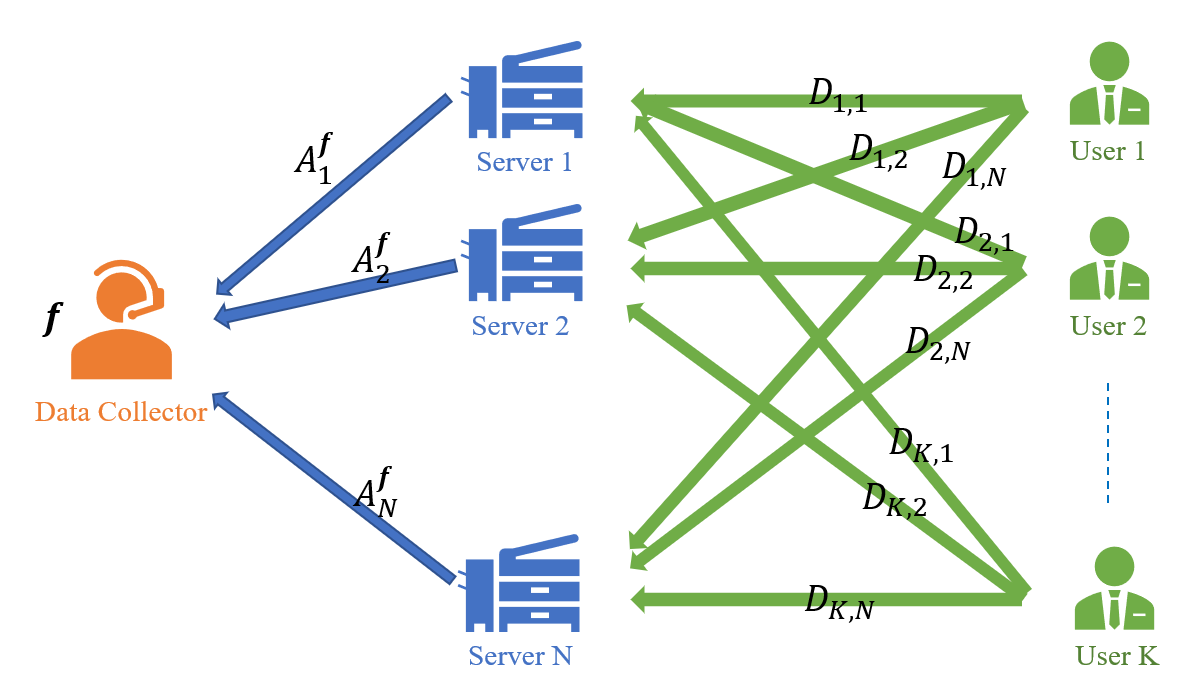}
	\caption{The Secure Privacy-preserving Epidemiological Data Collection Problem}
	\label{model}
\end{figure}

We formulate the secure privacy-preserving epidemiological data collection problem over a typical distributed secure computation system, in which there are  $K$ users, $N$ servers and a data collector, with their respective concerns on data security and privacy. 
The model is depicted in Fig. \ref{model}.

Assume that user $k$, $ {k\in[1:K]} $ has his/her personal message $ W_k $, and the messages of all users are independent and have an equal length of $ L $ symbols over a finite field $ GF_q $, i.e.,
\begin{align}
H(W_{[1:K]})=&\sum_{k=1}^{K}H(W_k),\\\quad\quad H(W_k)=&L,\quad \forall k\in [1:K].
\end{align}

The data collection problem contains two phases: in the first phase, which is called \emph{the upload phase}, all $ K $ users are required to upload a coded information to each of the $ N $ servers, where  the uploaded content  to the $ n $-th server by the $ k $-th user is denoted as $ D_{k,n}\in \mathfrak{D} $. The users would like to keep his/her message secure, more specifically, any up to $ E $ servers will learn nothing about the messages uploaded by the $K$ users, i.e., 
\begin{align}
I(D_{[K],\mathcal{E}};W_{[K]})=0,\quad \forall \mathcal{E}\subseteq[1:N],|\mathcal{E}|\leq E, \label{P1}
\end{align}
This is called \emph{the privacy constraint of the users against $E$ servers}. 
In order to achieve this, User $k$ utilizes a privately generated random noise $ Z_k \in\mathfrak{Z}_k$, i.e., $ Z_k $ is known to only User $k$, $k \in [1:K]$. The uploaded content of user $k$ is a deterministic function of $W_k$ and $Z_k$, i.e.,, there exists $ K $ functions $ d_k:GF_q^L\times \mathfrak{Z}_k\to \mathfrak{D}^N , k\in[1:K]$ that
$ d_k(W_k,Z_k)=\begin{bmatrix}
D_{k,1}\ D_{k,2},\cdots, D_{k,N}
\end{bmatrix}^T $, in other words,
\begin{align}
	H(D_{k,[1:N]}|W_k,Z_k)=0,\quad \forall k\in[1:K]. \label{221207a}
\end{align}

At the beginning of the second phase, which is called \emph{the computation phase}, a data collector would like to compute a statistics of the $ K $ messages of the users. In our setting, the statistics, denoted by $ W^{\mathbf{f}} $, is taken as a linear combination of all messages $ W_{[K]} $ with the coefficient $ \mathbf{f}\in GF_q^N $, i.e.,
\begin{align}
W^{\mathbf{f}}=f(W_{[K]})=\mathbf{f}^T\begin{bmatrix}
	W_1\\ \vdots \\ W_K
\end{bmatrix}= \sum_{k=1}^{K}f_kW_k,
\end{align}
and the value of $\mathbf{f}$ does not depend on the users' messages $W_{[1:K]}$. 
It is worth noticing that $\mathbf{f}$ is privately generated by the data collector, and furthermore, it is not known to the users and the servers during the upload phase. Hence, $D_{k,n}$ is not a function of $\mathbf{f}$ for all $k \in [1:K], n \in [1:N]$. 

In order to get the statistics $ W^{\mathbf{f}} $, the data collector generates designed queries to Server $n$, denoted as $Q_{n}^{\mathbf{f}} \in \mathfrak{Q}_n $, $n\in [1:N]$. Note that $Q_{n}^{\mathbf{f}}$ is not only a function of $\mathbf{f}$, but also a function of a randomnesses $ Z'\in\mathcal{Z} $, which is used to prevent $\mathbf{f}$ from being known by any single server. More specifically, the query $Q_{n}^{\mathbf{f}}$ is a deterministic function of $\mathbf{f}$ and $Z^\prime$, i.e., there exists a function $ q:GF_q^K\times \mathcal{Z}\to \prod_{n=1}^{N}\mathfrak{Q}_n $ that
\begin{align}
q(\mathbf{f},Z')=\begin{Bmatrix}
Q_1^{\mathbf{f}}\ Q_2^{\mathbf{f}}\ \cdots Q_N^{\mathbf{f}}
\end{Bmatrix}^T
\end{align}
Hence, we have
\begin{align}
	H(Q_{[1:N]}^{\mathbf{f}}|Z^\prime,\mathbf{f})=0, \label{221207b}
\end{align}
Since the data collector has no knowledge of the messages of the users nor the privately generated noise at the users, we have 
\begin{align}
	I(W_{[1:K]}, Z_{[1:K]};Q_{[1:N]}^{\mathbf{f}},Z^\prime)=0. \label{Nan08}
\end{align}

Upon receiving the query, all $ N $ servers are required to calculate the corresponding answers, denoted as $A_{[1:N]}^{\mathbf{f}}$, and send them to the data collector. More specifically, the answer generated by Server $n$, i.e., $A_n^{\mathbf{f}}\in \mathfrak{A}_n$, is a deterministic function of the stored content of Server $n$, i.e., $ D_{[1:K],n} $, and the query it received, i.e., $ Q_n^{\mathbf{f}} $. In other words, there exists the function
$
  a_n^{\mathbf{f}}:\mathfrak{Q}_n\times \mathfrak{D}^K\to \mathfrak{A}_n,\ A_n^{\mathbf{f}}=a_n^{\mathbf{f}}(Q_n^{\mathbf{f}},\begin{bmatrix}
  D_{1,n}\ D_{2,n},\cdots, D_{K,n}
  \end{bmatrix}^T),\quad n\in[1:N],
$
in other words, we have
\begin{align}
	H(A_n^{\mathbf{f}}|Q_n^{\mathbf{f}},D_{[1:K],n})=0, \quad\forall n\in[1:N]. \label{Nan11}
\end{align}

We would like to design the queries of the data collector to meet the following 3 constraints. The first constraint requires that the data collector is able to reconstruct the desired statistics from all the answers from the $ N $ servers, which we call \emph{the decodability constraint}. Let $ \phi $ be the reconstruction function of the data collector, where
$\phi:\prod_{n=1}^{N}\mathfrak{A}_n\times \prod_{n=1}^{N}\mathfrak{Q}_n\times GF_q^K\times \mathcal{Z}\to GF_q^L$, 
and 
\begin{align}
\hat{W}^{\mathbf{f}}=\phi(A_{[1:N]}^{\mathbf{f}},Q_{[1:N]}^{\mathbf{f}},\mathbf{f}, Z').\label{ReFun}
\end{align}
The probability of decoding error is given by
\begin{align}
P_e= \max_{\mathbf{f}}\text{Pr}\{\hat{W}^{\mathbf{f}}\neq {W}^{\mathbf{f}}\}. 
\end{align}
According to Fano's inequality, the decodability constraint is equal to
\begin{align}
	H(W^{\mathbf{f}}|A_{[1:N]}^{\mathbf{f}},Q_{[1:N]}^{\mathbf{f}}, Z')=&o(L), \quad \forall \mathbf{f}\in GF_q^N, \label{CC}
\end{align}
when $P_e \rightarrow 0$. 

For the second constraint, the data collector must learn only the statistics and nothing more about the messages of the users, which we call \emph{the privacy constraint of the users against the data collector}, i.e.,
\begin{align}
	I(W_{[1:K]};A_{[1:N]}^{\mathbf{f}},Q_{[1:N]}^{\mathbf{f}},Z'|W^{\mathbf{f}})=&0. \label{P2}
\end{align}

We further assume that the servers are curious about the coefficients of the statistics, i.e., $ \mathbf{f} $. To protect the privacy of the data collector, the third constraint is that we require the coefficient vector $ \mathbf{f} $ is not leaked to any single server even when the server somehow obtained all the users' messages, i.e.,
\begin{align}
	(A_{n}^{\mathbf{f}},Q_{n}^{\mathbf{f}},W_{[1:K]}) \sim (A_{n}^{\mathbf{f}'},Q_{n}^{\mathbf{f}'},W_{[1:K]}), \quad \forall \mathbf{f}, \mathbf{f}' \text{ linear independent} \label{P3}
\end{align}
This is called \emph{the privacy constraint of the data collector against non-colluding servers}. 

The reason why in the upload phase, we consider up to $E$ servers may collude, and in the computation phase, we consider non-colluding servers is the following: the upload phase and the computation phase do not always occur at the same time. For example, the users are required to upload their epidemiological data on a regular basis, while the data collector may start his/her queries to a certain statistics at a relatively random time. Due to the dynamic topology of the servers, the numbers of colluding servers may be different during the uploading phase and the computation phase. Our work assumes that the servers are non-colluding in the computation phase, since the servers may be more interested in the epidemiological data. If $E=1$, then we have a model where the servers are non-colluding in both the upload phase and the computation phase. 

For any scheme that satisfies the above decodability constraint, i.e., (\ref{CC}), and the privacy constraints, i.e., the privacy constraint of the users against $E$ servers (\ref{P1}), the privacy constraint of the users against the data collector (\ref{P2}), and the privacy constraint of the data collector against the non-colluding servers (\ref{P3}), its communication rate is characterized by the number of symbols the data collector decodes per download symbol, i.e.,
\begin{align}
	R:=\frac{L}{\sum_{n=1}^{N}H(A_{n}^\mathbf{f})}. 
\end{align}
Note that $R$ is not a function of $\mathbf{f}$ due to (\ref{P3}).

A rate $ R $ is said to be ($ \epsilon $-error) achievable if there exists a sequence of schemes with their communication rate less than or equal to $ R $ where the probability of error $ P_e $ goes to zero as $ L\to\infty $. The $ \epsilon $-error capacity of this random secure aggregation problem is defined as the supremum of all $ \epsilon $-error achievable rates, i.e., $ C :=\sup R $, where the supremum is over all possible $ \epsilon $-error achievable schemes. 

\section{Main Result}
\begin{Theo}\label{main1}
	When the number of users $ K\to \infty $, the asymptotic capacity of the secure privacy-preserving epidemiological data collection problem is
\begin{align}
	\lim\limits_{K\to\infty, L\to\infty}C=\begin{cases}\frac{N-E-1}{N}, \quad &\text{ if } E< N-1\\
		0, \quad &\text{ otherwise }
	\end{cases},
\end{align}
\end{Theo}

The converse proof of Theorem \ref{main1} will be given in Section \ref{Con}, and the achievability proof for certain cases of finite $ K\in\mathbb{N}_+ $ will be given in Section \ref{Achi}. Noticing that when $ K\to \infty $, the schemes in Section \ref{Achi} can be achievable by sending multiple rounds of queries using the same strategy, the rate of achievability and converse will meet when $ K $ goes infinity.



\begin{Remark}
	When the number of users $ K $ is a finite integer, the achievability and converse results of our work do not meet. From our derivations, the converse for finite $ K $ seems to be depend on $ K $, while the scheme we construct is irrelevant to $ K $ in order to protect the privacy of the users against the data collector. How to close the gap when $ K $ is finite is still an open problem.
\end{Remark}

\section{Proof of Theorem \ref{main1}: Converse when $ E<N-1 $}\label{Con}
In this section, we prove the converse part of Theorem \ref{main1} when $E<N-1$. First, we prove the following lemma which states an iterative relationship on the number of linear combinations of the users' messages. 
\begin{Lem}
	Let $ \mathbf{f}_1,\mathbf{f}_2, \cdots, \mathbf{f}_K \in GF_q^K $ be linear independent vectors, and $ \mathcal{E}\subseteq [1:N] $, $ |\mathcal{E}|=E $, we have
	\begin{align}
	&H(A_{[1:N]/\mathcal{E}}^{\mathbf{f}_{k}}|W^{\mathbf{f}_1},\cdots, W^{\mathbf{f}_{k}},D_{[1:K],\mathcal{E}},Q_{[1:N]}^{\mathbf{f}_k}, Z') \nonumber\\
	& \geq  \frac{L}{N-E}+\frac{1}{N-E}H(A_{[1:N]/\mathcal{E}}^{\mathbf{f}_{k+1}}|W^{\mathbf{f}_1},\cdots, W^{\mathbf{f}_{k+1}},D_{[1:K],\mathcal{E}}, Q_{[1:N]}^{\mathbf{f}_{k+1}}, Z')-o(L)\label{ConIte}
\end{align}
\end{Lem}
\begin{IEEEproof}
	\begin{align}
	&(N-E)H(A_{[1:N]/\mathcal{E}}^{\mathbf{f}_k}|W^{\mathbf{f}_1},\cdots, W^{\mathbf{f}_k}, D_{[1:K],\mathcal{E}},Q_{[1:N]}^{\mathbf{f}_k}, Z')\nonumber\\
	\geq &\sum_{n\in [1:N]/\mathcal{E} }H(A_{n}^{\mathbf{f}_k}|W^{\mathbf{f}_1},\cdots, W^{\mathbf{f}_k}, D_{[1:K],\mathcal{E}},Q_{[1:N]}^{\mathbf{f}_k},Z')\label{ConIne2}\\
	=&\sum_{n\in [1:N]/\mathcal{E} }H(A_{n}^{\mathbf{f}_{k+1}}|W^{\mathbf{f}_1},\cdots, W^{\mathbf{f}_k}, D_{[1:K],\mathcal{E}},Q_{[1:N]}^{\mathbf{f}_{k+1}},Z') \label{Nan15}\\
	\geq& H(A_{[1:N]/\mathcal{E}}^{\mathbf{f}^\prime}|W^{\mathbf{f}},D_{[1:K],\mathcal{E}}, Q_{[1:N]}^{\mathbf{f}^\prime}, Z') \nonumber\\%
	=&H(A_{[1:N]/\mathcal{E}}^{\mathbf{f}^\prime}|W^{\mathbf{f}},D_{[1:K],\mathcal{E}}, Q_{[1:N]}^{\mathbf{f}^\prime}, Z')+H(W^{\mathbf{f}^\prime}|A_{[1:N]/\mathcal{E}}^{\mathbf{f}^\prime}, W^{\mathbf{f}},D_{[1:K],\mathcal{E}},Q_{[1:N]}^{\mathbf{f}^\prime}, Z')-o(L)\label{Nan16}\\
	=& H(W^{\mathbf{f}^\prime}|W^{\mathbf{f}},D_{[1:K],\mathcal{E}},Q_{[1:N]}^{\mathbf{f}^\prime})+H(A_{[1:N]/\mathcal{E}}^{\mathbf{f}^\prime}|W^{\mathbf{f}^\prime},W^{\mathbf{f}},D_{[1:K],\mathcal{E}},Q_{[1:N]}^{\mathbf{f}^\prime}, Z' )-o(L)\nonumber \\
	=&L+H(A_{[1:N]/\mathcal{E}}^{\mathbf{f}^\prime}|W^{\mathbf{f}^\prime},W^{\mathbf{f}},D_{[1:K],\mathcal{E}},Q_{[1:N]}^{\mathbf{f}^\prime}, Z')-o(L) \label{Nan17}
\end{align}
where (\ref{ConIne2}) holds because of the non-negativity of $ H(A_{[1:N]/(\mathcal{E}\cup\{k\})}^{\mathbf{f}_k}|A_{n}^{\mathbf{f}_k},W^{\mathbf{f}_1},\cdots, W^{\mathbf{f}_k}, D_{[1:K],\mathcal{E}},Q_{[1:N]}^{\mathbf{f}_k},Z') $ for all $ n\in [1:N]/\mathcal{E} $, (\ref{Nan15}) holds because of (\ref{P3}). The equality in (\ref{Nan16}) follows due to the fact that $H(W^{\mathbf{f}^\prime}|A_{[1:N]/\mathcal{E}}^{\mathbf{f}^\prime}, W^{\mathbf{f}},D_{[1:K],\mathcal{E}},Q_{[1:N]}^{\mathbf{f}^\prime}, Z')=H(W^{\mathbf{f}^\prime}|A_{[1:N]/\mathcal{E}}^{\mathbf{f}^\prime}, W^{\mathbf{f}},D_{[1:K],\mathcal{E}}, A_{\mathcal{E}}^{\mathbf{f}^\prime},Q_{[1:N]}^{\mathbf{f}^\prime}, Z')=o(L)$, where the first equality follows from (\ref{Nan11}), and the second equality follows from (\ref{CC}). Finally, (\ref{Nan17}) holds because $ W^{\mathbf{f}^\prime} $ is independent from the queries and randomness, the security constraint and that $ \mathbf{f},\mathbf{f}^\prime\in GF_q^K $ are linear independent vectors.
\end{IEEEproof}

The following lemma shows that any set of answers are independent from any queries conditioning on the same set of queries to the same coefficient and any size of messages and randomnesses. This is the direct inference from the independence of message, queries and randomnesses generated by the data collector (\ref{Nan08}).
\begin{Lem} \label{Nan01}
	Assume that $ \mathbf{f}\in GF_q^N $, $ \mathcal{N}_1,\mathcal{N}_2\in[1:N] $, and $ \mathcal{K}\in[1:K] $, we have the following equality:
	\begin{align}
	I(A_{\mathcal{N}_1}^{\mathbf{f}};Q_{\mathcal{N}_2}^{\mathbf{f}}|W_{\mathcal{K}},Z^\prime,Q_{\mathcal{N}_1}^{\mathbf{f}})
\end{align}
\end{Lem}
\begin{IEEEproof}
	The proof is the same as \cite[Section VI, Lemma 1]{wang2018capacity}, and the key of this proof is that $ A_{\mathcal{N}_1}^{\mathbf{f}} $ is determined by $ W_{[1:K]} $, conditioning on $ Z^\prime$ and $Q_{\mathcal{N}_1}^{\mathbf{f}} $. We omit the detailed proof here.
\end{IEEEproof}

The lemma below has a similar form of Lemma \ref{Nan01}, and it shows that any set of answers with size of $ E $ do not dependent on the desired statistic, conditioning on the same set of queries and the randomnesses generated by the data collector. 

\begin{Lem}\label{Nan02}
	For any $ \mathcal{E}\subseteq[1:N], |\mathcal{E}|=E $,
	\begin{align}
	H(A_{\mathcal{E}}^{\mathbf{f}}|Q_{\mathcal{E}}^{\mathbf{f}},W^{\mathbf{f}},Z^\prime)=&H(A_{\mathcal{E}}^{\mathbf{f}}|Q_{\mathcal{E}}^{\mathbf{f}},Z^\prime)\label{Nan02Eq}
\end{align}
\end{Lem}
\begin{IEEEproof}
	we only need to show that $ I(A_{\mathcal{E}}^{\mathbf{f}};W^{\mathbf{f}}|Q_{\mathcal{E}}^{\mathbf{f}},Z^\prime) $ is less than or equal to $ 0 $ because of its non-negativity.
	\begin{align}
		I(A_{\mathcal{E}}^{\mathbf{f}};W^{\mathbf{f}}|Q_{\mathcal{E}}^{\mathbf{f}},Z^\prime)
		\leq&I(A_{\mathcal{E}}^{\mathbf{f}},D_{[1:K],\mathcal{E}};W^{\mathbf{f}}|Q_{\mathcal{E}}^{\mathbf{f}},Z^\prime)\\
		=&I(D_{[1:K],\mathcal{E}};W^{\mathbf{f}}|Q_{\mathcal{E}}^{\mathbf{f}},Z^\prime)+I(A_{\mathcal{E}}^{\mathbf{f}};W^{\mathbf{f}}|D_{[1:K],\mathcal{E}},Q_{\mathcal{E}}^{\mathbf{f}},Z^\prime)\\
		=&I(D_{[1:K],\mathcal{E}};W^{\mathbf{f}}|Q_{\mathcal{E}}^{\mathbf{f}},Z^\prime)\label{Nan02P1}\\
		=&H(D_{[1:K],\mathcal{E}}|Q_{\mathcal{E}}^{\mathbf{f}},Z^\prime)-H(D_{[1:K],\mathcal{E}}|W^{\mathbf{f}},Q_{\mathcal{E}}^{\mathbf{f}},Z^\prime)\\
		=&0\label{Nan02P2}
	\end{align}
where \ref{Nan02P1} holds because the answers $ A_{\mathcal{E}}^{\mathbf{f}} $ are determined by $ (D_{[1:K],\mathcal{E}},Q_{\mathcal{E}}^{\mathbf{f}},Z^\prime) $ in (\ref{Nan11}), and (\ref{Nan02P2}) holds because of (\ref{Nan08}) and (\ref{P1}).
\end{IEEEproof}

The following lemma shows that we can split the answers into two parts, one from $ E $ servers that cannot decode the database and the other from $ N-E $ servers:
\begin{Lem}\label{Nan04}
	For any $ \mathbf{f}\in GF_q^K $ and $ \mathcal{E}\in[1:N] $, $ |\mathcal{E}|=E $, we have
	\begin{align}
	&\left(1-\frac{E}{N}\right)H(A_{[1:N]}^{\mathbf{f}}|Q_{[1:N]}^{\mathbf{f}},Z^\prime)
	\geq L+H(A_{[1:N]/\mathcal{E}}^{\mathbf{f}}|W^{\mathbf{f}},D_{[1:K],\mathcal{E}},Q_{[1:N]}^{\mathbf{f}},Z^\prime)-o(L)\label{ConIni}
\end{align}
\end{Lem}
\begin{IEEEproof}
Based on the system model, we have
	\begin{align}
	H(A_{[1:N]}^{\mathbf{f}}|Q_{[1:N]}^{\mathbf{f}},  Z')=&H(W^{\mathbf{f}}|Q_{[1:N]}^{\mathbf{f}},  Z')+H(A_{[1:N]}^{\mathbf{f}}|W^{\mathbf{f}},Q_{[1:N]}^{\mathbf{f}},  Z')-H(W^{\mathbf{f}}|A_{[1:N]}^{\mathbf{f}},Q_{[1:N]}^{\mathbf{f}}, Z') \nonumber\\
	=&L+H(A_{[1:N]}^{\mathbf{f}}|W^{\mathbf{f}},Q_{[1:N]}^{\mathbf{f}},  Z')-o(L) \label{Nan07}\\
	=&L+H(A_{\mathcal{E}}^{\mathbf{f}}|W^{\mathbf{f}},Q_{[1:N]}^{\mathbf{f}}, Z')+H(A_{[1:N]/\mathcal{E}}^{\mathbf{f}}|W^{\mathbf{f}},A_{\mathcal{E}}^{\mathbf{f}},Q_{[1:N]}^{\mathbf{f}}, Z')-o(L) \nonumber\\%
	=&L+H(A_{\mathcal{E}}^{\mathbf{f}}|W^{\mathbf{f}},Q_{\mathcal{E}}^{\mathbf{f}}, Z')+H(A_{[1:N]/\mathcal{E}}^{\mathbf{f}}|W^{\mathbf{f}},A_{\mathcal{E}}^{\mathbf{f}},Q_{[1:N]}^{\mathbf{f}}, Z')-o(L) \label{Nan09}\\
	=&L+H(A_{\mathcal{E}}^{\mathbf{f}}|Q_{\mathcal{E}}^{\mathbf{f}}, Z')+H(A_{[1:N]/\mathcal{E}}^{\mathbf{f}}|W^{\mathbf{f}},A_{\mathcal{E}}^{\mathbf{f}},Q_{[1:N]}^{\mathbf{f}}, Z')-o(L) \label{Nan10}\\%
	\geq&L+H(A_{\mathcal{E}}^{\mathbf{f}}|Q_{\mathcal{E}}^{\mathbf{f}}, Z') +H(A_{[1:N]/\mathcal{E}}^{\mathbf{f}}|W^{\mathbf{f}},D_{[1:K],\mathcal{E}},Q_{[1:N]}^{\mathbf{f}},  Z')-o(L) \label{Nan12}\\
\geq &L+\frac{E}{N}H(A_{[1:N]}^{\mathbf{f}}|Q_{[1:N]}^{\mathbf{f}})+H(A_{[1:N]/\mathcal{E}}^{\mathbf{f}}|W^{\mathbf{f}},D_{[1:K],\mathcal{E}},S,Q_{[1:N]}^{\mathbf{f}}, Z')-o(L)\label{Nan13}
\end{align}
\end{IEEEproof}
where (\ref{Nan07}) follows from (\ref{Nan08}) and (\ref{CC}), (\ref{Nan09}) follows from Lemma \ref{Nan01} when $ \mathcal{N}_1=\mathcal{E} $, $ \mathcal{N}_2=[1:N] $,  (\ref{Nan10}) follows from (\ref{Nan02Eq}), (\ref{Nan08}) and (\ref{Nan11}), (\ref{Nan12}) is because of (\ref{Nan11}), and (\ref{Nan13}) follows from the Han's inequality,
\begin{align}
	\sum_{\mathcal{E}\subseteq[1:N], |\mathcal{E}|=E}H(A_{\mathcal{E}}^{\mathbf{f}}|Q_{\mathcal{E}}^{\mathbf{f}})\geq \frac{E}{N}\binom{N}{E}H(A_{[1:N]}^{\mathbf{f}}|Q_{[1:N]}^{\mathbf{f}})
\end{align}

Now, we can get the lower bound on the asymptotic download size when $ L $ and $ K $ goes infinity as 
\begin{align}
	&\lim\limits_{K\to\infty,L\to\infty}\left(1-\frac{E}{N}\right)H(A_{[1:N]}^{\mathbf{f}}|Q_{[1:N]}^{\mathbf{f}})\nonumber \\
	\geq &\lim\limits_{K\to\infty,L\to\infty} \left(H(A_{[1:N]/\mathcal{E}}^{\mathbf{f}}|W^{\mathbf{f}},D_{[1:K],\mathcal{E}},S,Q_{[1:N]}^{\mathbf{f}})+L-o(L)\right)\label{Nan05}\\
	= &\lim\limits_{K\to\infty,L\to\infty} H(A_{[1:N]/\mathcal{E}}^{\mathbf{f}}|W^{\mathbf{f}},D_{[1:K],\mathcal{E}},S,Q_{[1:N]}^{\mathbf{f}})+L\label{Nan06}\\
	\geq &\lim\limits_{K\to\infty,L\to\infty} \frac{1}{N-E}\left(L+H(A_{[1:N]/\mathcal{E}}^{\mathbf{f}^\prime}|W^{\mathbf{f}}, W^{\mathbf{f}'}, D_{[1:K],\mathcal{E}},S,Q_{[1:N]}^{\mathbf{f}^\prime}-o(L))\right)+L\nonumber\\
	\geq& \left(\sum_{k=0}^{\infty}\frac{1}{(N-E)^k}\right)L\nonumber
\end{align}
where (\ref{Nan05}) follows from (\ref{ConIni}), and (\ref{Nan06}) is because $ o(L) $ goes zero when $ L\to \infty $. Thus we can calculate the upper bound of the asymptotic capacity when $ E<N-1 $ as follows
\begin{align}
	\lim\limits_{K\to\infty, L\to\infty}C \leq &\lim\limits_{K\to\infty, L\to\infty}\frac{L}{H(A_{[1:N]}^{\mathbf{f}})}\nonumber\\
	\leq&\lim\limits_{K\to\infty, L\to\infty} \frac{L}{H(A_{[1:N]}^{\mathbf{f}}|H(Q_{[1:N]}^{\mathbf{f}})}\nonumber\\
	\leq &\frac{1-\frac{E}{N}}{\sum_{k=0}^{\infty}\frac{1}{(N-E)^k}}\nonumber\\
	=& \frac{N-E-1}{N}\nonumber
\end{align}

\section{Proof of Theorem \ref{main1}: Achievability when $ E<N-1 $}\label{Achi}

In this section, we give a cross subspace alignment (CSA) scheme based on the coding of interference in the computation phase to reach the asymptotic capacity\cite{Jia2019XSTPIR} for any integer $ N> E+1$ and $K\geq 2 $. Throughout the scheme, we choose the length of each personal message $ L=N-E-1\geq 1 $, and we use the notation $ \Delta_n=\prod_{i=1}^{L}(i+\alpha_n) $ for $ n\in[1:N] $.

First, we specify the encoding functions $ \{d_k\}_{k\in [1:K]} $ in the upload phase. Let $ W_k^l\in GF_q $ be the $ l$-th symbol of each $ W_{k} $, $ k\in[1:K], l \in [1:L] $ and $ W^l\in GF_q^{1\times K}$ be the row vector of the $l$-th symbol of all $K$ messages, i.e., $W^l=[W_1^l,\cdots,W_K^l] $. Assume that $ \alpha_n, n\in[1:N] $ are $N$ distinct coefficients all belonging to the set $\{\alpha\in GF_q:\alpha +i\neq 0, i\in[1:L] \} $,  i.e., for any $ i,j\in[1:N], \alpha_1\neq \alpha_j $. Note that the $\alpha_n$s are globally shared variables, known to the users, servers and the data collector. In order to protect the privacy of the users against the servers, each user $k$ will generate $L \times E$ random noises $Z_{le}^k$ uniformly from $GF_q$.
The uploaded information to the $n$-th server by the $k$-th user is given by
\begin{align}
	D_{k,n}
	=\begin{bmatrix}
		W_k^1+\sum_{e=1}^{E}(1+\alpha_n)^eZ_{1e}^k \\
		\vdots \\ W_k^L+\sum_{e=1}^{E}(L+\alpha_n)^eZ_{Le}^k
	\end{bmatrix}^T\forall k\in[1:K],n\in[1:N],\label{DStorage1}
\end{align}


For notational convenience, we write the content stored at Server $n$ in a vector form as 
\begin{align}
	D_n=&[D_{1,n}^{1},\cdots,D_{K,n}^{1},D_{1,n}^{2},\cdots,D_{K,n}^{2},\cdots,D_{1,n}^{L},\cdots,D_{K,n}^{L}]\\
	=&\begin{bmatrix}
		W^1+\sum_{e=1}^{E}(1+\alpha_n)^eZ_{1e} \\
		\vdots \\ W^L+\sum_{e=1}^{E}(L+\alpha_n)^eZ_{Le}.
	\end{bmatrix}^T
\end{align}
where $ D_n\in GF_q^{1\times KL} $, and $Z_{le}$ is defined as $Z_{le}=[Z_{le}^1, \cdots, Z_{le}^K]$.



In the computation phase, the query to Server $ n $ is determined by the coefficient $ \mathbf{f} $ and the randomness from data collector $ Z^\prime $.
We design the query to Server $n$ based on $\mathbf{f}$ as
\begin{align}
	Q_{n}^{\mathbf{f}}
	=\begin{bmatrix}
		\frac{\Delta_n}{1+\alpha_n}(\mathbf{f}+(1+\alpha_n)Z^\prime_{1})\\
		\vdots\\
		\frac{\Delta_n}{L+\alpha_n}(\mathbf{f}+(L+\alpha_n)Z^\prime_{L})
	\end{bmatrix}
\end{align}
where $Z_1',\cdots, Z_{L}'$ are $L$ random column vectors of length $K$, whose elements are uniformly distributed on $GF_q$, generated by the data collector. 

 For any server $ n\in[1:N] $, the answer to the data collector $ A_{n}^{\mathbf{f}}\in \mathfrak{A}_n=GF_q $ is calculated by 
\begin{align}
	&A_{n}^{\mathbf{f}}=D_n\cdot Q_{n}^{\mathbf{f}}\\
	=& \left(W^1+\sum_{e=1}^{E}(1+\alpha_n)^eZ_{1e}\right)\cdot\left(\frac{\Delta_n}{1+\alpha_n}(\mathbf{f}+(1+\alpha_n)Z^\prime_{1})\right)\nonumber\\
	&+\cdots+\left(W^L+\sum_{e=1}^{E}(L+\alpha_n)^eZ_{Le}\right)\cdot\left(\frac{\Delta_n}{L+\alpha_n}(\mathbf{f}+(L+\alpha_n)Z^\prime_{L})\right),\forall n\in[1:N].\label{B1}
\end{align}
As can be seen, $\frac{A_{n}^{\mathbf{f}}}{\Delta_n}$ is the sum of $\sum_{l=1}^{L}\frac{1}{l+\alpha_n}W^l\cdot\mathbf{f}$ and a polynomial of degree $E$ in $\alpha_n$, whose coefficient in front of $\alpha_n^e$ is denoted as $I_e$, $e \in [0:E]$. Note that $I_e$ is not a function of $n$. 

%
Writing the answers to the data collector from the servers in a matrix form, we have
\begin{align}
	\begin{bmatrix}
		\frac{A_1^{\mathbf{f}}}{\Delta_1}\\\frac{A_2^{\mathbf{f}}}{\Delta_2}\\\cdots \\\frac{A_N^{\mathbf{f}}}{\Delta_N}
	\end{bmatrix}
	=\begin{bmatrix}
		\frac{1}{1+\alpha_1}\ \cdots \ \frac{1}{L+\alpha_1} \ 1 \ \alpha_1 \ \cdots \ \alpha_1^E\\
		\frac{1}{1+\alpha_2}\ \cdots \ \frac{1}{L+\alpha_2} \ 1 \ \alpha_2 \ \cdots \ \alpha_2^E\\
		\cdots \\
		\frac{1}{1+\alpha_N}\ \cdots \ \frac{1}{L+\alpha_N} \ 1 \ \alpha_N \ \cdots \ \alpha_N^E\\
	\end{bmatrix}\cdot
	\begin{bmatrix}
		W^1\cdot\mathbf{f}\\
		\vdots\\
		W^L\cdot\mathbf{f}\\
		I_{0}\\
		\vdots\\
		I_{E}
	\end{bmatrix}, \label{221207c}
\end{align}

Now we prove that this scheme satisfies the decodability constraint, i.e., (\ref{CC}), and the privacy constraints, i.e., the privacy constraint of the users against $E$ servers (\ref{P1}), the privacy constraint of the users against the data collector (\ref{P2}), and the privacy constraint of the data collector against the non-colluding servers (\ref{P3}). 

Recall that in our scheme, we let $L=N-E-1$. The decodability constraint is satisfied because the matrix in (\ref{221207c}) is a $ N\times N $ full-rank matrix when the $ \alpha_n $s are distinct\cite[Lemma 5]{Jia2019XSTPIR}. Hence, $W^1\cdot\mathbf{f}, \cdots, W^L\cdot\mathbf{f}$ may be recovered from $\frac{A_1^{\mathbf{f}}}{\Delta_1}, \cdots,\frac{A_N^{\mathbf{f}}}{\Delta_N}$, and as we know 
\begin{align}
	W^{\mathbf{f}}=W^T\cdot\mathbf{f}
	=\begin{bmatrix}
		W^1\cdot\mathbf{f}\\
		\cdots\\
		W^L\cdot\mathbf{f}
	\end{bmatrix}. 
\end{align}

%
%
%

The privacy constraint of the users against $E$ colluding servers is satisfied
due to the sharing strategy of the users. 
In (\ref{DStorage1}), we know that the $ k $-th user share its $ l $-th symbol to the $ n $-th server in a form 
\begin{align}
	D_{k,n}^l= W_k^l+\sum_{e=1}^{E}(1+\alpha_n)^eZ_{le}^k\label{Storage},
\end{align}
where $ D_{k,n}^l $ denotes the storages in server $ n $ that $ W_k^l $ shares. The security need to guarantee that any $ E $ out of $ N $ servers do not know $ W_k^l $ for any $ k\in[1:K] $ and $ l\in[1:L] $. As the storage in each server has the same form (\ref{Storage}), we can choose the $ E $ servers to be in $ [1:E] $, w.l.o.g. We write the storage of these servers with respect to what $ W_k^l $ shares in a matrix form
\begin{align}
	\begin{bmatrix}
		D_{k,1}^l \\
		\cdots \\
		D_{k,E}^l
	\end{bmatrix}=&
	\begin{bmatrix}
		W_k^l \\
		\cdots \\
		W_k^l
	\end{bmatrix}+
	\begin{bmatrix}
		l+\alpha_1 \ (l+\alpha_1)^2 \ \cdots \ (l+\alpha_1)^E \\
		\cdots \\
		l+\alpha_E \ (l+\alpha_1)^E \ \cdots \ (l+\alpha_E)^E \\
	\end{bmatrix}\cdot
	 \begin{bmatrix}
		Z_{l1}^k \\
		\cdots \\
		Z_{lE}^k
	\end{bmatrix}.\label{St2}
\end{align}

Notice that the Vandermonde matrix in (\ref{St2}), denoted by $ V_E $, is invertible for distinct $ \{1+\alpha_e:\alpha_e\in GF_q, e\in[1:E]\} $, so the second term of (\ref{St2}) contains $ E $ symbols that are linearly independent, and we have can prove the privacy constraint against $ E $ colluding servers that 
\begin{align}
	&I(D_{k,\mathcal{E}};W_{k})\nonumber\\
	\leq&\sum_{i=1}^{L}\sum_{j=1}^{L}I(D_{k,\mathcal{E}}^i;W_{k}^j|D_{k,\mathcal{E}}^{[1:i-1]};W_{k}^{[1:j-1]})\nonumber\\
	=&\sum_{i=1}^{L}\sum_{j=1}^{L}I(W_k^i\mathbf{1}_{E}+V_E\cdot Z_{i,\mathcal{E}}^k;W^{j}|D_{[1:i-1],\mathcal{E}};W^{[1:j-1]})\label{PE01}\\ =&\sum_{l=1}^{L}I\left(W_k^l;W_k^l\mathbf{1}_E+V_E\cdot Z_{l,\mathcal{E}}^k\right)\label{PE02}\\
	=&\sum_{l=1}^{L}I\left(W_k^l;
		Z_{l,\mathcal{E}}^k\right)\nonumber\\
		=&0,\quad\quad \forall k\in[1:K],n\in[1:N].\nonumber
\end{align}
where (\ref{PE01}) comes from (\ref{St2}), and (\ref{PE02}) holds because that when $ i<j $, we have
\begin{align}
&I(W_k^i\mathbf{1}_{E}+V_E\cdot Z_{i,\mathcal{E}}^k;W^{j}|D_{[1:i-1],\mathcal{E}};W^{[1:j-1]})\nonumber\\
=& H(W_k^i\mathbf{1}_{E}+V_E\cdot Z_{i,\mathcal{E}}^k|D_{[1:i-1],\mathcal{E}};W^{[1:j-1]})-H(W_k^i\mathbf{1}_{E}+V_E\cdot Z_{i,\mathcal{E}}^k|D_{[1:i-1],\mathcal{E}};W^{[1:j]})\nonumber\\
=&H(W_k^i\mathbf{1}_{E}+V_E\cdot Z_{i,\mathcal{E}}^k|W^i)-H(W_k^i\mathbf{1}_{E}+V_E\cdot Z_{i,\mathcal{E}}^k|W^i)=0\nonumber
\end{align}
and when $ i>j $, we have
\begin{align}
&I(W_k^i\mathbf{1}_{E}+V_E\cdot Z_{i,\mathcal{E}}^k;W^{j}|D_{[1:i-1],\mathcal{E}};W^{[1:j-1]})\nonumber\\
=& H(W_k^i\mathbf{1}_{E}+V_E\cdot Z_{i,\mathcal{E}}^k|D_{[1:i-1],\mathcal{E}};W^{[1:j-1]})-H(W_k^i\mathbf{1}_{E}+V_E\cdot Z_{i,\mathcal{E}}^k|D_{[1:i-1],\mathcal{E}};W^{[1:j]})\nonumber\\
=&H(W_k^i\mathbf{1}_{E}+V_E\cdot Z_{i,\mathcal{E}}^k)-H(W_k^i\mathbf{1}_{E}+V_E\cdot Z_{i,\mathcal{E}}^k)=0\nonumber
\end{align}
so the remaining items are those satisfying $ i=j=l $.

To prove that the privacy constraint of the data collector against the non-colluding servers is satisfied, 
we notice that the query to each server is composed of the desired coefficient $ \mathbf{f} $ and independent and identically distributed additional noise $ Z_l^\prime $, $ l\in[1:L] $. Thus, any single server can not distinguish queries from data collector with one coefficient $ \mathbf{f} $. To be specific, we have
\begin{align}
	&I(Q_{n}^{\mathbf{f}},A_{n}^{\mathbf{f}},W_{[1:K]};\mathbf{f})\nonumber\\
	\leq& I(Q_{n}^{\mathbf{f}},W_{[1:K]},Z;\mathbf{f})\label{P11}\\
	=&I(Q_{n}^{\mathbf{f}};\mathbf{f}|W_{[1:K]},Z)\label{P12}\\
	\leq& H(Q_{n}^{\mathbf{f}})-H\left(\left.\begin{bmatrix}
		\mathbf{f}+(1+\alpha_n)(Z^\prime_{1})\\
		\cdots\\
		\mathbf{f}+(L+\alpha_n)(Z^\prime_{L})
	\end{bmatrix}\right|\mathbf{f},W_{[1:K]},Z\right)\nonumber\\
	=&H(Q_{n}^{\mathbf{f}})-H(Z^\prime_{1},\cdots,Z^\prime_{L})\nonumber\\
	=&L-L\nonumber\\
	=&0,\nonumber
\end{align}
where (\ref{P11}) is from (\ref{B1}), and (\ref{P12}) is from (\ref{P2}).

Finally, to prove that the privacy constraint of the users against the data collector is satisfied, we construct a basis of $ GF_q^K $ containing the desired $ \mathbf{f} $. Assume that the vectors in the basis is denoted by $ \{\mathbf{f}_1,\mathbf{f}_2,\cdots,\mathbf{f}_{K}\} $ where $ \mathbf{f}_1=\mathbf{f} $. We then have
\begin{align}
	&I\left(W_{[1:K]};A_{[1:N]}^{\mathbf{f}}, Q_{[1:N]}^{\mathbf{f}}, Z'|W^{\mathbf{f}}\right) \nonumber\\
	=&\sum_{l=1}^{L}I\left(W^{l};A_{[1:N]}^{\mathbf{f}},Q_{[1:N]}^{\mathbf{f}}, Z'|W^{[1:l-1]},W^{\mathbf{f}}\right) \nonumber\\
	\leq&\sum_{l=1}^{L}I\left(W^{l};A_{[1:N]}^{\mathbf{f}},Q_{[1:N]}^{\mathbf{f}}, Z'|W^{[1:L]/\{l\}},Z,W^{\mathbf{f}}\right)\label{S11}\\
	=&\sum_{l=1}^{L}I\left(\left(W^l+\sum_{e=1}^{E}(l+\alpha_n)^eZ_{le}\right)\cdot\right.\nonumber\\
	&\quad\quad\quad\quad \left.\left(\frac{\Delta_n}{1+\alpha_n}(\mathbf{f}+(l+\alpha_n)(Z^\prime_{l})^T)\right)_{n\in[1:N]};W^l|W^{[1:L]/\{l\}},Z,W^{\mathbf{f}}\right)\label{S12}\\
	=&\sum_{l=1}^{L}I\left(\left( W^l(Z^\prime_{l})^T+\sum_{e=1}^{E}(l+\alpha_n)^eZ_{le}(Z^\prime_{l})^T     \right)_{n\in[1:N]};W^l|W^{[1:L]/\{l\}},W^{\mathbf{f}}\right)\label{S13}\\
	\leq &\sum_{l=1}^{L}I\left(\left( W^l(Z^\prime_{l})^T,Z_{le}(Z^\prime_{l})^T     \right)_{n\in[1:N],e\in[1:E]};W^l|W^{[1:L]/\{l\}},W^{\mathbf{f}}\right)\label{S14}\\
	=&0,
\end{align}
where (\ref{S11}) holds because $ (W^{[1:L]/\{l\}},Z) $ is independent with $ W^l $, (\ref{S12}) holds because except for the term containing $ W^l $, all terms in (\ref{B1}) are given, so deducting them will not change the mutual information. (\ref{S13}) is because $ \Delta_n $ is a constant.

Thus, we can prove that the scheme satisfies all the constraints. As any server answer is a symbol from $ GF_q $, the rate in the proposed scheme is
\begin{align}
R=\frac{L}{H(A_n^{\mathbf{f}})}=\frac{N-E-1}{N}
\end{align}

We notice that the achievable rate meets the asymptotic upper bound for any $ K\in\mathbb{N}_+ $, so the scheme is then proved to be asymptotically optimal, by letting $ K\to\infty $.

\section{proof of theorem \ref{main1}: converse when $ E\geq N-1 $ }
We know that when $ N=E $, the correctness and security constraints contradicts each other and any scheme is not feasible to the problem, so we focus on the case when $ N=E+1 $.

For converse, the inequality (\ref{ConIni}) also holds, as the inequality in (\ref{ConIne2}) becomes an equality, and we have
\begin{align}
	H(A_{[1:N]/\mathcal{E}}^{\mathbf{f}}|W^{\mathbf{f}},D_{[1:K],\mathcal{E}},S,Q_{[1:N]}^{\mathbf{f}},Z^\prime) 
	\geq& L+H(A_{[1:N]/\mathcal{E}}^{\mathbf{f}^\prime}|W^{\mathbf{f}},W^{\mathbf{f}^\prime},D_{[1:K],\mathcal{E}},Q_{[1:N]}^{\mathbf{f}^\prime},Z^\prime)-o(L)\label{NE01}
\end{align}
so for any linear independent vectors $ \mathbf{f}_1,\mathbf{f}_2, \cdots, \mathbf{f}_K \in GF_q^K $, we have 
\begin{align}
	\left(1-\frac{E}{N}\right)H(A_{[1:N]}^{\mathbf{f}}|Q_{[1:N]}^{\mathbf{f}}) \geq & H(A_{[1:N]/\mathcal{E}}^{\mathbf{f}_1}|W^{\mathbf{f}_1},D_{[1:K],\mathcal{E}},Q_{[1:N]}^{\mathbf{f}_1},Z^\prime)+L-o(L)\label{NE02}\\
	\geq &  2L+H(A_{[1:N]/\mathcal{E}}^{\mathbf{f}_2}|W^{\mathbf{f}_1},D_{[1:K],\mathcal{E}},Q_{[1:N]}^{\mathbf{f}_2},Z^\prime)-o(L)\label{NE03}\\
	\geq& \sum_{k=0}^{K}L+H(A_{[1:N]/\mathcal{E}}^{\mathbf{f}_K}|W^{[1:L]},D_{[1:K],\mathcal{E}},Q_{[1:N]}^{\mathbf{f}_K},Z^\prime)\label{NE04}\\
	=&\sum_{k=0}^{K}L+H(A_{[1:N]/\mathcal{E}}^{\mathbf{f}_K}|W_{[1:K]},D_{[1:K],\mathcal{E}},Q_{[1:N]}^{\mathbf{f}_K},Z^\prime)\label{NE05}\\
	=&KL\nonumber
\end{align}
where (\ref{NE02}),(\ref{NE03}) and (\ref{NE04}) are from (\ref{NE01}), and (\ref{NE05}) holds because any $ K $ linear independent vectors constitute a basis in $ GF_q^K $, so $ W_{[1:K]} $ can be decoded. We can see that the download will go infinite when $ K\to\infty $, and thus the asymptotic capacity will be
\begin{align}
	\lim\limits_{K\to\infty, L\to\infty}C=&\lim\limits_{K\to\infty, L\to\infty}\frac{L}{H(A_{[1:N]}^{\mathbf{f}})}\\
	\leq& \lim\limits_{K\to\infty, L\to\infty}\frac{L}{H(A_{[1:N]}^{\mathbf{f}}|H(Q_{[1:N]}^{\mathbf{f}})}\\
	\leq &\lim\limits_{K\to\infty, L\to\infty}\frac{1-\frac{E}{N}}{KL}\\
	=& 0
\end{align}

The upper bound of $ C $ indicates that it is unfeasible to construct a scheme that has a positive communication rate when $ K\to \infty $.
\section{Conclusion}
We have modeled and found the asymptotical capacity of the privacy-preserving epidemiological data collection problem. We show that when there are more than 1 remaining servers that do not collude with other servers to decode the users' data, the asymptotical capacity exists. The results in this work shows a similar capacity form with symmetric private information retrieval.

\bibliographystyle{unsrt}
\bibliography{main}

\end{document}